# Wetting dynamics on lyophilic solid surfaces patterned by lyophobic islands

Roumen Tsekov[1], Dimitar Borissov[2] and Stoyan I. Karakashev[1]
[1]Department of Physical Chemistry, University of Sofia, 1164 Sofia, Bulgaria
[2]Nanosemiconductor GmbH, Konrad-Adenauer-Allee 11, 44263 Dortmund, Germany

A theory for wetting of structured solid surfaces is developed, based on the delta-comb periodic potential. It possesses two matching parameters: the effective line tension and the friction coefficient on the three-phase contact line at the surface. The theory is validated on the dynamics of spreading of liquid zinc droplets on morphologically patterned zinkophilic iron surface by means of square patterns of zinkophobic aluminum oxide. It is found out that the effective line tension is negative and has essential contribution to the dynamics of spreading. Thus, the theoretical analysis shows that the presence of lyophobic patterns situated on lyophilic surface makes the latter completely wettable, i.e. no equilibrium contact angle on such surface exists making the droplet spread completely in form of thin liquid layer on the patterned surface.

Wetting as phenomenon plays a crucial role in a number industrial processes such as mineral flotation, antifoam performance, spraying of paints, metallurgical coatings, tertiary oil recovery, inkjet spray coatings, dentistry, etc. However, the poor control on wetting in some technological processes causes often engineering problems. Thus, for example, to achieve a high quality hot-dip galvanization of steel sheets, a good wetting between the latter and the melted zinc is needed [1]. Unfortunately however, the oxides cover on the steel surface is non-wettable by the melted zinc resulting in poor quality of the final product [2, 3]. To remove the oxides, the steel sheets are usually subjected to series of annealing and re-crystallization in forming gas atmosphere prior the hot-dip galvanization [4]. Yet, some oxides as $Al_2O_3$ and $SiO_2$ cannot be eliminated. Such surface usually is covered with wettable and non-wettable by the liquid zinc fragments. Moreover, the oxidized fragments are morphologically more elevated as compared to the non-oxidized part of the surface. Understanding the physics of wetting of such kind of surface is necessary to improve the performance of the hot dip galvanization.

When some liquid and solid get in contact, they allow the adhesive, cohesive, and capillary forces to counter-balance each other thus determining the properties of the three-phase contact line (TPCL) and the very regime of dynamic wetting/dewetting [5-9]. It has been found that the lyophobicity [9] and morphology [10-13] of the solid substrate affect essentially the regime of wetting. To study these effects, particular methods for chemical and morphological [14] pattering of solid substrates were developed. It was established [10, 15-19] that when a liquid drop gets in contact with energetically heterogeneous surface a number of meta-stable configurations caused by the surface heterogeneity and the vibrational energy of the liquid drop control the contact angle hysteresis. Yet, the latter is related to the line tension of the droplet [20, 21]. Moreover, it was reported [22-24] that the line tension plays an important role only in the cases of small liquid droplets in contact with a solid surface. The droplet shapes on a chemically and morphologically inhomogeneous surface were studied in the literature as well [24-28]. It was shown that the chemically heterogeneous surface causes wetting transitions at which the shape of the wetting edge changes in a specific and usually abrupt manner. If the droplet is substantially larger than the surface patterns the contact angle usually is described by the classical Cassie-Baxter equation. In a recent paper [29], we have developed a theory on motion of the three-phase contact (TPCL) of evaporating very small droplet situated on chemically patterned smooth glass surface with alternating lyophobic and lyophilic areas. For this reason, the delta-comb potential was used on modeling the surface energy distribution of such kind solid surface. We use hereafter the same theoretical approach to develop a theory of spreading very small liquid zinc droplet on lyophilic iron Fe sheet covered with alternating lyophobic micro-squares of Aluminum oxide $Al_2O_3$ with a height in the nano-scale. Similarly to the hot-dip galvanization, the iron sheet is wetted very well by the melted zinc, while the alternating squares from $Al_2O_3$ on its surface exhibits poor wetting. We aim with this work at shedding a light on this industrial problem from theoretical viewpoint.

The experiment details are already published [2, 3]. For wettability study, an iron sheet with a thickness of 2 mm (Goodfellow, 99.95%) was cut into pieces of 15 mm x 15 mm. The specimens were wet ground with SiC papers (400 to 2000 grit) and followed by polishing down to 1 μm with diamond paste on nylon cloth, afterwards degreased in ethanol and dried under a stream of pure nitrogen. Patterning was carried out by fixing via four magnets a squared aluminum mesh with windows of 130 μm x 130 μm and a periodicity of 210 μm. The specimens were introduced to a Physical Vapor Deposition (PVD) chamber (Univex 450, Leybold) and α-$Al_2O_3$ (99.99%, Merck) was evaporated with a thickness of 20 nm. During the vapor deposition, a pressure of about $2 \times 10^{-5}$ mbar kept constant. The deposition rate was 1 Å/s and the thickness of the $Al_2O_3$ patterns were controlled in-situ by quartz crystal microbalance (QCM, Inficon). The wettability by liquid zinc of aluminum oxide patterned surfaces was investigated by sessile-drop method. The controlled wetting experiments were conducted in a home built liquid zinc spin coater setup. The setup is ultra-high-vacuum (UHV) compatible and contains a load lock, main chamber and a preparation chamber. The main chamber atmosphere consisted of a mixture of $N_2$-5%$H_2$ with a dew point of -70°C. This low dew point was required to avoid the oxidation of the Zn droplet before it is deposited upon the surface. The main chamber was constantly purged with a flow rate of 5 slm. The aluminum oxide patterned specimens were fixed to a sample holder which could be resistance heated up to 1000 °C and then were introduced into the load lock, which was evacuated and backfilled with $N_2$-5%$H_2$ gas atmosphere of dew point -70 °C. Afterwards, the specimens were transferred to the main chamber where they were annealed for 1 minute at 820 °C in $N_2$-5%$H_2$. The temperature was measured by a K-type thermocouple at the front of the sample. At the top of the main chamber, a syringe is mounted to deposit a droplet of liquid zinc upon the sample surface. In the syringe, the zinc is molten in a quartz glass capillary embedded in a graphite body. The temperature of the melt was maintained at 460°C. The bath was composed of a Zn-0.2w%Al alloy which is generally used as galvanizing bath. All samples were heated at 6 °C/s. After annealing, the samples were cooled to 470°C in one minute. During this time, the samples were approached from 3 to 5 mm under the syringe. When the sample temperature was stabilized at 470°C, a single zinc droplet was deposited on the surface. The sample was kept at 470°C for 3 minutes before final cooling to room temperature. The wetting behavior of the zinc droplet was monitored by a video system, which afterwards allowed measuring the contact angle as a function of the wetting time (see Fig.1).

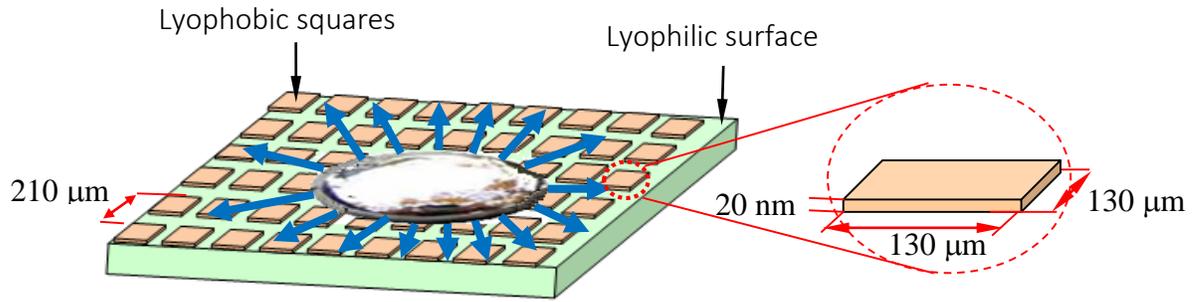

**Fig. 1** Sketch on spreading of liquid zinc droplet on the patterned surfaces (not to scale)

First, we would like to preliminary stipulate the basic terminology in this chapter and in the whole paper as well. As far as this work explores the dynamics of spreading of liquid zinc on iron substrate patterned with $Al_2O_3$ islands, the terms lyophilic and lyophobic regard the affinity of the solid substrate with respect to the liquid zinc. Yet, despite the natures of the liquid and the substrate, the laws of wetting are universal. As was mentioned in the introduction, the problem of wetting of structured surfaces possesses both equilibrium and kinetic aspects. For instance, the equilibrium, advancing and receding contact angles could be strongly affected by the Cassie or Wenzel states. The present paper is trying to describe the wetting by liquid Zn of a Fe surface patterned by square $Al_2O_3$ islands, spread periodically with a lattice parameter $a$. Let us consider a small drop of liquid Zn deposited on this structured solid surface, where the drop radius is much larger than the characteristic size of an $Al_2O_3$ island. Since the weight of the drop is 25 mg and the density of the liquid Zn at 460°C is 6.86 g/cm³ [30], the drop volume amounts to $V = 3.64$ µL. If the radius of the three-phase contact line of the drop is denoted by $r$, at relatively slow wetting its evolution in time obeys the following dynamic equation

$$b\, dr/dt = \sigma_{SG} - \sigma_{SL} - \sigma_{LG} \cos\theta \qquad (1)$$

where the first term is the friction force with $b$ being the TPCL friction coefficient. On the right hand side $\sigma_{SG}$, $\sigma_{SL}$ and $\sigma_{LG}$ are the solid/gas, solid/liquid and liquid/gas surface tensions, respectively, while $\theta$ is the local non-equilibrium contact angle. On a homogeneous $Al_2O_3$ surface the three surface tensions above define via the Young-Laplace equation an equilibrium contact angle $\theta_{Al_2O_3}$, which is about 140°. On the structured surface an additional force per unite length $\bar{\Delta}$ acting on the contact line should be added to obtain

$$\bar{\Delta} = \sigma_{SG} - \sigma_{SL} - \sigma_{LG} \cos\theta_{Al_2O_3} \tag{2}$$

The parameter $\bar{\Delta}$ determines the local equilibrium contact angle and could take into account either Cassie or Wenzel states.

To model the effect of the remaining Fe stripes square lattice one can employ a delta-comb potential, which corresponds to the following effective potential energy per unite surface [29]

$$\Delta(x,y) = -\sum_j \pi\kappa[\delta(x-ja)/2 + \delta(y-ja)/2] \tag{3}$$

where $\delta(\cdot)$ is the Dirac delta-function, $a$ is the lattice parameter and $\kappa$ is a specific constant accounting for the interfacial energy differences on Fe and $Al_2O_3$ surfaces. Since the drop possess radial symmetry one can calculate the angular average of this energy as follows

$$\bar{\Delta}(r) = \frac{1}{2\pi}\int_0^{2\pi} \Delta(r\cos\varphi, r\sin\varphi)d\varphi \approx \kappa[\arcsin(a/r)/a - \pi/2a] \approx (2a\kappa - \pi\kappa r)/2ar \tag{4}$$

The last simplified expression here takes into account the fact that the drop is much larger than the pattern characteristic size, i.e. $r \gg a$. Introducing now Eq. (4) in Eq. (2) and substituting the result in Eq. (1) yields

$$bdr/dt = \sigma_{LG}(\cos\theta_{Al_2O_3} - \cos\theta) - \pi\kappa/2a + \kappa/r \tag{5}$$

It is evident from Eq. (5) that the parameter $\kappa$ represents, in fact, an effective line tension on the drop three-phase contact. It determines also the equilibrium contact angle, if there is any, on the structured surface, $\cos\theta_e = \cos\theta_{Al_2O_3} - \pi\kappa/2a\sigma_{LG}$, which corresponds to a constant $r$. Since the Fe and $Al_2O_3$ patterns possess different lyophobicity (the contact angle of pure Fe surface is about 6°) and also a different height (the $Al_2O_3$ islands are about 20 nm higher) this equilibrium angle accounts for both the Cassie and Wenzel effects. To complete Eq. (5) another relation is required between the three-phase contact line radius and contact angle. Assuming nearly hemi-spherical shape of the drop the drop volume $V$ can be expressed as a function of the three-phase contact radius and contact angle

$$V = \frac{\pi r^3}{3} \frac{2+\cos\theta}{1+\cos\theta} \sqrt{\frac{1-\cos\theta}{1+\cos\theta}} \qquad (6)$$

Assuming that the droplet volume is constant because evaporation is negligible, Eq. (6) provided the necessary relation between $r$ and $\cos\theta$. Thus, the system of Eqs. (5) and (6) describes the evolution of all geometric parameters of the drop. It can be integrated numerically and compared to experimental observation to determine the two unknown parameters, $b$ and $\kappa$, while the surface tension of Zn at 460 °C is $\sigma_{LG} = 0.754$ N/m [25].

The experimental details are already published [1, 2]. For wettability study, a polished iron sheet with size 15 x 15 x 2 mm was patterned by a squared α-Al$_2$O$_3$ mesh with windows of 100 x 100 μm and a periodicity of 180 μm. The specimen with a thickness of 20 nm was introduced by a physical vapor deposition. The wettability by liquid zinc of these patterned surfaces was investigated by sessile-drop method. The controlled wetting experiments were conducted in a home built liquid zinc spin coater UHV setup.

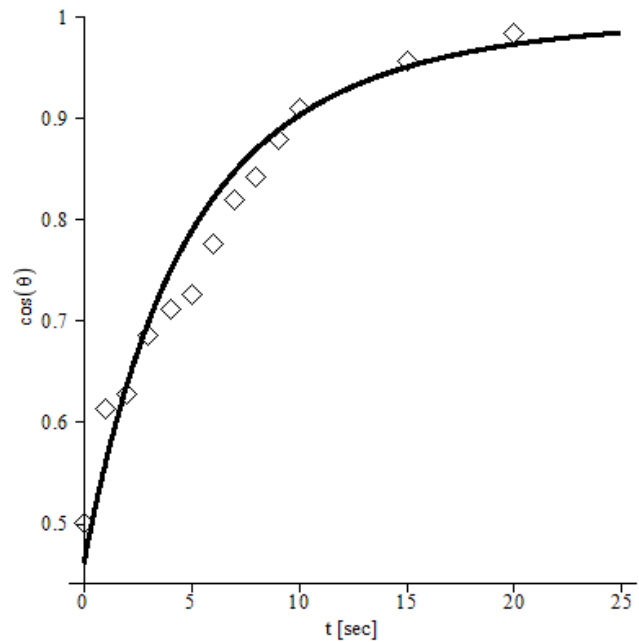

**Fig. 2** Experimental points and theoretical fit of the evolution of the drop $\cos\theta$

In Fig. 2 the experimental points of $\cos\theta$ vs. $t$ are fitted by numerical solution of Eqs. (5) and (6). The determined value of the parameters are $b = 1$ Ns/cm² and $\kappa = -0.22$ mN. The negative val-ue of the effective line tension is due to the lyophilic nature of the Fe stripes. Since the value $\cos\theta_{Al_2O_3} - \pi\kappa/2a\sigma_{LG} = 1.8$ is larger than 1 the structured surface is completely wettable and there is no equilibrium contact angle. Hence, the Wenzel effect makes the lyophilic Fe surface a completely wettable one. The radius of the three-phase contact is simultaneously calculated by $\cos\theta$ and plotted in Fig. 3. As is seen initially it amounts to about 8 surface structural units but during the evolution $r$ riches linearly a value more than $15 \times a = 2.7$ mm. The latter corresponds well to the experimental observations. It is well-known that the surface roughness, despite being natural or fabricated, makes the lyophilic surface more lyophilic.

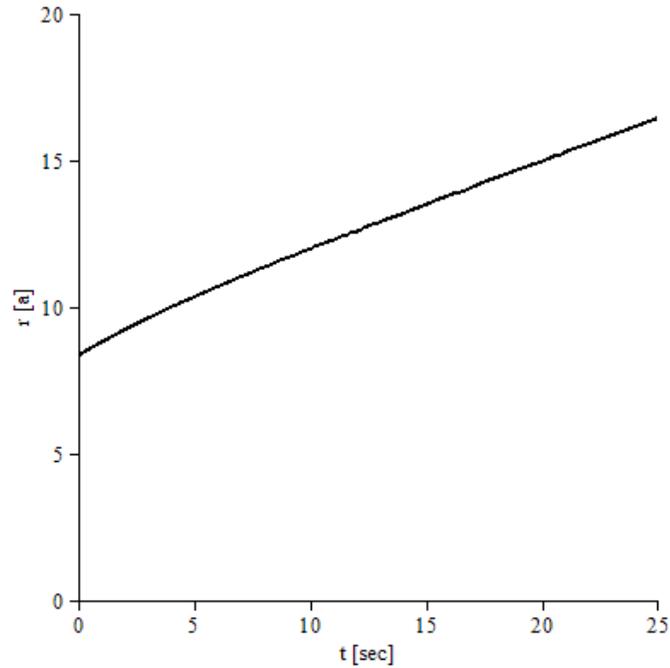

**Fig. 3** Calculated evolution of the radius of the three-phase contact $r$ expressed in $a$ units

The present work reports about a case when a periodical surface roughness in form of lyophobic pillars on lyophilic surface makes the last one completely wettable. Under such condition, the droplet situated on the surface keeps spreading until reaching ultra-thin surface layer. The latter as phenomenon can have substantial practical applications in the coating industry. Moreover, Eq. (5) is based on the presumption that the radius of the droplet is substantially larger than the parameter of periodicity ($r \gg a$). One can see in Fig. 3 that the ratio $r/a$ increases upon time in the range 8-17. The present work studies the kinetics of spreading of liquid zinc (Zn) on morphologically patterned iron (Fe) surface with lyophobic pillars of aluminum oxide ($Al_2O_3$). A general theory about such class of phenomena was developed. The theory contains two matching parameters: the effective line tension and the friction coefficient of TPCL on patterned surface. It was found out a negative value of the line tension contributing for complete spreading of the liquid on the surface. Thus the lyophilic surface patterned with lyophobic pillars became completely wettable one. The latter as fundamental point is important due to the fact that it is well known that the surface roughness makes the hydrophobic surface super-hydrophobic one due to the Cassie state. Now it is reported that surface roughness can make the lyophilic surface completely wettable one due to the Wenzel effect. In our opinion both chemical heterogeneity and surface roughness are responsible for this effect, although the chemical heterogeneity have stronger contribution in making the value of line tension negative.